# An Order Optimal Policy for Exploiting Idle Spectrum in Cognitive Radio Networks

Jan Oksanen, *Student Member, IEEE* and Visa Koivunen, *Fellow, IEEE*

*Abstract*—In this paper a spectrum sensing policy employing recency-based exploration is proposed for cognitive radio networks. We formulate the problem of finding a spectrum sensing policy for multi-band dynamic spectrum access as a stochastic restless multi-armed bandit problem with stationary unknown reward distributions. In cognitive radio networks the multi-armed bandit problem arises when deciding where in the radio spectrum to look for idle frequencies that could be efficiently exploited for data transmission. We consider two models for the dynamics of the frequency bands: 1) the independent model where the state of the band evolves randomly independently from the past and 2) the Gilbert-Elliot model, where the states evolve according to a 2-state Markov chain. It is shown that in these conditions the proposed sensing policy attains asymptotically logarithmic weak regret. The policy proposed in this paper is an index policy, in which the index of a frequency band is comprised of a sample mean term and a recency-based exploration bonus term. The sample mean promotes spectrum exploitation whereas the exploration bonus encourages for further exploration for idle bands providing high data rates. The proposed recency based approach readily allows constructing the exploration bonus such that it will grow the time interval between consecutive sensing time instants of a suboptimal band exponentially, which then leads to logarithmically increasing weak regret. Simulation results confirming logarithmic weak regret are presented and it is found that the proposed policy provides often improved performance at low complexity over other state-of-the-art policies in the literature.

*Index Terms*—Cognitive radio, opportunistic spectrum access (OSA), restless multi-armed bandit (RMAB), online learning, multi-band spectrum sensing

## I. INTRODUCTION

When looking at the radio spectrum allocation tables one might come to the discouraging conclusion that there are very few radio resources available for new wireless systems and services. However, a very different conclusion is reached when one actually measures the true utilization of the radio spectrum at a particular location and time. In fact it has been demonstrated by measurement campaigns (e.g. [1]) that many parts of the spectrum are heavily underutilized. Cognitive radio (CR) is a technology that holds promise for more efficient use of such underutilized radio spectrum. A cognitive radio network consists of secondary users (SUs) that sense the spectrum for idle frequencies that they could use for transmission in an agile manner. When the SUs sense that a part of the spectrum allocated to the primary users (PUs) is idle, the SUs may use those frequencies for data transmission. When the PUs become active again the SUs need to be able to detect the primary signal and vacate the band and search for idle frequencies elsewhere in the spectrum. For an up-to-date and extensive review of different spectrum sensing techniques as well as various exploration and exploitation schemes for CR, see, for example [2], [3].

Depending on the location and the spectral allocation of the PUs, some frequency bands are idle more often than others. Some bands may have higher bandwidths and experience less interference, thus, potentially support higher data rates. Naturally, it is desirable that the CRs focus on sensing those bands that are persistently available and have a large bandwidth, i.e., bands that are expected to provide high data rates. However, since the expected data rates are in practice unknown, the CR needs to learn them. This learning problem resembles a variation of the multi-armed bandit (MAB) [4] problem where the objective is to learn and identify the subbands that provide the highest data rates while not wasting too much time on sensing frequencies with low data rates. In machine learning this is also referred to as the *exploration-exploitation* tradeoff problem [5].

In the classical MAB problem a player is faced with $K$ slot machines (one armed bandits) the $k^{\text{th}}$ one of which produces an unknown expected reward $\mu_k$, $k = 1, ..., K$. The player's goal is to collect as much reward as possible over time, i.e., to learn the machine that has the highest $\mu_k$ as fast as possible. The dynamic rule governing the selection of the machine to be played is called a policy and it is commonly denoted by $\pi$. In cognitive radio the analogous counterpart of a slot machine is a frequency band suitable for wireless transmissions whereas a reward corresponds to an achieved data rate when the secondary user accesses an idle band. A good recent review of the classical bandit problem and its variations is given in [4].

The general description of the MAB problem given above contains a rich blend of variations with different assumptions and solutions. The MAB problems in the literature may be broadly categorized into problems with independent rewards and problems with Markovian rewards. These two categories may be further subdivided into rested and restless bandit problems and furthermore into problems with known statistics and problems with unknown statistics. In the independent MAB the rewards of each machine are generated by a time independent random process whereas in the Markovian MAB the rewards are generated by a Markov chain. The difference between the rested MAB and the restless MAB (RMAB) is that in the rested MAB the state of the machine can change only when the machine is played, while in the restless MAB the

The authors are with the Department of Signal Processing and Acoustics, Aalto University, FI-00076 AALTO, Finland (e-mail: jan.oksanen@aalto.fi, visa.koivunen@aalto.fi).

state of the machine keeps evolving regardless of whether it is played or not. The restlessness of the MAB has implications only in the Markov case but not in the independent reward case, since independent rewards naturally do not depend on the past rewards nor the players actions.

The optimal strategy for the Markovian rested MAB with known statistics is the so called Gittins index policy [6]. The MAB with Markovian restless rewards with known statistics is in general PSPACE-hard [7], but for a particular relaxed version of the problem the optimal solution has been provided by Whittle [8]. The optimal policy for the rested and restless MAB problem with time independent rewards is obviously to always play the machine with the highest expected reward. For the restless Markovian MAB problem with unknown statistics the optimal policy is generally not known.

In this paper we are interested in the RMAB problem with unknown statistics - a problem that stems from multi-channel dynamic spectrum access in cognitive radio. In particular we consider the case where the rewards are independent in time and the case where they evolve according to a 2-state Markov chain (i.e. the Gilbert-Elliot model). The 2-state model captures the fact that the spectrum is either idle or occupied, hence it is particularly suitable for the CR problem. In the RMAB problem the states of the non-activated machines may change, similarly as the state of the spectrum band may change regardless of whether it is sensed or not. Hence the RMAB is a suitable model for the dynamic spectrum access problem. Depending for example on the lengths of the time slots of the SU and the PU either the independent reward model or the Markovian reward model may be more appropriate. For example, if the operational time slot of the secondary user is much smaller than the time slot of the primary user the Markovian model may be more suitable. This is because in that case the PUs consecutive actions may be correlated from the SUs point of view. On the other hand if the SU has a much larger time slot than the primary user the independent reward model can be more appropriate. In this paper we cover both independent and Markovian reward cases.

In the Markov case, since computing the optimal policy is in general PSPACE-hard [7], a weaker notion of optimality has been used in the literature called *the best single arm policy* [9], [10], [11]. The best single arm policy is defined as the policy that produces the highest cumulative reward by always playing only one arm, which is the arm with the highest stationary mean reward. Note that in the i.i.d. MAB the best single arm policy is also the optimal policy. In the rest of this paper the term optimal policy always refers to the best single arm policy.

The success of a policy $\pi$ is measured by its expected weak regret which is the difference between the expected total payoff using policy $\pi$ and the total payoff expected when the best single arm policy $\pi^*$ is used. In [12] it was shown that when the rewards are independent for any policy, the weak regret is asymptotically lower bounded by a function that grows logarithmically in time. Consequently, policies achieving logarithmic weak regret are called order optimal.

This paper proposes a sensing policy that is asymptotically order optimal when the rewards are independent. Furthermore, it is shown that asymptotically logarithmic expected weak regret is achieved when the rewards are restless and follow a Markov chain with two states (i.e. the Gilbert-Elliot model). The proposed policy is an index policy consisting of a sample mean term and an exploration bonus term. The sample mean promotes exploitation whereas the exploration bonus encourages for exploration. The exploration bonus in this paper is based on *recency*, i.e., it promotes exploring such bands that have not been sensed for a long time. The higher the exploration bonus of a particular band, the more likely the band will be explored in the near future. The exploration bonus is designed such that asymptotically the time difference of two consecutive sensing time instants on a suboptimal band grows exponentially, which also provides an effortless intuition for logarithmic weak regret. In this paper we in fact show that the proposed policies achieve asymptotic logarithmic weak regret and demonstrate by simulations that they can often outperform other state-of-the-art policies or obtain equal performance with reduced complexity. Asymptotic results are typically achieved with finite sample size. However, the point at which the asymptotic result begins to hold need to be determined by simulations. Another advantage of the recency based policies proposed in this paper is that the tradeoff between exploration and exploitation becomes asymptotically deterministic. This allows for simplifying the proposed policy in a practical implementation. For instance, in centralized cooperative spectrum sensing, where a fusion center (FC) maintains and runs the sensing policy on behalf of the (possibly unintelligent) SUs, one would like to minimize the amount of control information transmitted between the FC and the SUs. After a sufficient number of sensings the exploration and exploitation time instances have practically become deterministic. Then the FC does not need to instruct the SUs at every time instant which band to sense. Instead, the FC needs to communicate to the SUs only at those time instants when the sensed band changes. This could significantly reduce the amount of control traffic the FC needs to transmit. However, we leave these kinds of developments and quantitative results for future studies.

The proposed recency-based policy may find applications also outside the spectrum sensing context. These possible areas of application are (but not limited to) adaptive clinical trials, webpage content experiments, internet advertising, game playing and learning online the shortest path in a graph with stochastic edge weights (see e.g. [13] and references therein).

*A. Contributions and structure of the paper*

Some preliminary ideas and results of this paper were published by the authors in [14], where a special case of the sensing policy proposed in this paper was developed. The contributions of this paper are the following:

- We generalise the idea of recency-based exploration in RMAB formulation of multi-band spectrum sensing and show how to find order optimal sensing policies for different reward (data rate) distribution families. In particular we find order optimal policies for bounded i.i.d. rewards and for rewards generated by the Gilbert-Elliot model.
- It is shown that the performance of the proposed sensing policy can be enhanced when the type of the reward

distribution is known, e.g., when the rewards are i.i.d. Bernoulli, by simply modifying a constant in the exploration bonus.
- A nontrivial analysis of the expected weak regret of the proposed policy for i.i.d. and Markovian rewards is provided and the weak regret is shown to be asymptotically logarithmic.
- We present extensive computer simulations demonstrating logarithmic weak regret of the proposed policy and demonstrate that the policy often outperforms other state of the art policies or achieves equal performance at significantly lower computational cost.

The rest of the paper is organized as follows. In Section II we give an overview of the related work in the area of MAB problems and sensing policies in dynamic spectrum access. In Section III we express mathematically using the RMAB formulation the problem of finding a spectrum sensing policy. In section IV we propose the spectrum sensing policy based on the RMAB formulation for both i.i.d. and for Markovian data rates (rewards). In section V we show how to optimize the exploration bonus for particular reward distribution classes. Section VI illustrates the performance of the policy and verifies the analytical results using simulation examples. The paper is concluded in Section VII.

## II. RELATED WORK

Since the seminal paper by Lai and Robbins [12], much of the work on the stochastic MAB problems has focused on index policies with low expected weak regret and low computational complexity. Many of these policies are built on a principle called "optimism in the face of uncertainty". This principle states that an agent (learner) should stay optimistic about actions whose exact expected response is uncertain. Policies based on this principle may be categorized in two groups: optimistic initial value policies [5], [15] and exploration bonus policies [12], [14], [15], [16], [17], [18].

In optimistic initial value policies the value of an action is initialized with a high bias in order to guarantee sufficient amount of exploration in the beginning of the learning process. In [15] it was shown that setting the initial value sufficiently high guarantees convergence to an $\epsilon$-optimal policy. However, selecting high enough initial values that lead also to a good finite time performance is not a trivial task, which makes these policies impractical for the purposes of this paper.

Policies based on exploration bonuses assign a bonus for the actions based on, for example confidence, frequency or recency. Confidence based policies [16], [12], [17] evoke optimism through the use of an optimistic upper confidence bound on the expected reward estimate of the actions which effectively makes insufficiently explored actions more attractive for the agent. Frequency based policies assign bonuses to the actions based on the number of times they have been taken. In this regard most confidence based policies, such as the UCB by [16], may also be seen to be frequency based since the value of the confidence bound is inversely proportional to the number of times the action has been taken. Recency-based policies, such as the one proposed in this paper, promote exploration in proportion to the time that has passed since the action was last tried. As a consequence, actions that have not been taken for a long time will be chosen more likely in the near future. The rate at which exploration is promoted is gradually decreased in time in order to guarantee convergence to the optimal action. To the best of our knowledge, this paper is the first to develop and analyze recency-based exploration in the bandit setting.

For the classical stochastic MAB problem with independent rewards it was shown in [12] (and later generalized in [19]) that asymptotically order optimal policies have an expected weak regret of $O(\log(n))$. In [20] this lower bound was further generalized for the case where the rewards are at rest and evolve according to an irreducible and aperiodic Markov chain. However, for general restless Markovian rewards (apart from the special case of i.i.d. rewards, such as Bernoulli) theoretical lower bounds on the weak regret have not been reported in the literature.

In [12] a class of confidence bound based policies that achieve asymptotical logarithmic weak regret was presented for the MAB problem. However, these policies require storing the entire history of the observed rewards, which makes their implementation impractical. The recency based policies proposed in this paper require storing only a sample mean term and an exploration bonus term for each of the frequency bands. In [21], a class of policies based on sample means and upper confidence bounds was proposed. These policies were simpler compared to those in [12]. However, the policies in [21] are distribution dependent and deriving the upper confidence bounds in a closed form is often tedious. Among the most celebrated bandit papers is [16], where a computationally simple upper confidence bound (UCB) policy was proposed and shown to be uniformly order optimal when the rewards are independent and have a bounded support. This policy was further developed in [22] (see Theorem 2.2 therein) by improving a constant in the UCB policy. The recency based policy proposed in this paper has similar desirable properties as the UCB policy in terms of its simple implementation. Additionally, the recency based policy has an intuitive explanation for its asymptotically logarithmic weak regret. In the UCB policy exploration bonus is based on confidence, whereas in this paper the policies are based on recency. In addition to this fundamental difference, we have observed in our simulations that the policy proposed in this paper often achieves lower weak regret than the UCB. Recently the KL-UCB policy was proposed in [23]. It is an asymptotically optimal policy (i.e. it achieves the lower bound of [12]) for bounded i.i.d. rewards whose distributions are known except for their parameterization. The KL-UCB was analytically shown to have uniformly lower expected weak regret than the UCB when the rewards are Bernoulli. Therefore, we compare the policy proposed in this paper to the KL-UCB policy instead of the UCB. However, the KL-UCB is computationally more expensive than the policy proposed in this paper (as well as the UCB) as it requires solving a constrained optimization problem using dichotomic search or Newton's iterations. For example the average time complexity of binary search is $O(\log_2(m))$, where $m$ is the size of the search grid that essentially sets the accuracy of the optimal upper confidence bound. In the proposed sensing policy no such search steps are required. Interestingly, although being much simpler, our simulations with independent rewards show that the proposed recency based policy performs equally well or close to the KL-UCB in many scenarios.

CR spectrum sensing policies stemming from the RMAB formulation have been proposed for example in [9], [10], [11], [24], [25], [26]. Among them [9] and [10] are the most relevant ones for this work since they also model the sensing problem as an RMAB problem with unknown statistics and use the weak regret as the performance measure. In [9] a policy achieving uniformly logarithmic weak regret in the RMAB problem with unknown Markovian rewards was proposed for centralized and decentralized CR networks. The policy proposed in [9] operates in regenerative cycles that cleverly allows for using an UCB type index policy. In this paper we also employ regenerative cycles when learning the stationary expectations of the Gilbert-Elliot channel in section IV-B. The policy in [9], however, discards all observations made outside the regenerative cycles making it an inefficient learner in some cases. In this paper however all collected observations are used for learning. In [10] a policy based on deterministic sequencing of exploration and exploitation (DSEE) epochs was proposed for the RMAB problem and shown to achieve uniform logarithmic weak regret in centralized and distributed channel sensing. The principle used in the policy proposed in this paper, which is to grow the periods of exploitation exponentially in time, echoes similar ideas as the policy in [10]. However, the policy proposed in this paper is an index policy whereas the DSEE policy operates explicitly in epochs of exploration and exploitation by maintaining a dynamic list of frequency bands to be either explored or exploited. The policy proposed in this paper has a simple index form. Moreover, our simulation results demonstrate that also a better performance is often obtained by using the proposed policy. The simulations also indicate that the deterministic nature of the DSEE exploration epochs occasionally results in sudden increases in the weak regret, which do not occur in the policy proposed in this paper. In [24] it was shown that under the assumption that the channels are identical and independent 2-state Markovian channels whose signs of correlation are known, the RMAB problem can be solved with a simple myopic sensing policy. However, in practice the channels would rarely obey the same statistics (e.g. radios are in different locations, scattering environments and experience different SINR and are mobile). Furthermore, the sign of correlation of the channels is usually not known a priori. In our paper the statistics of the underlying rewards are not assumed to be known. In [25] the problem of finding optimal sensing policy was cast as a partially observable Markov decision process (POMDP) with unknown channels' state transition probabilities. The proposed algorithm works by estimating the transition probabilities during exploration phases and then mapping the obtained estimates in to so called policy zones for which the optimal policies have been precomputed. However, apart from a few special cases, it is not always possible to precompute the optimal policies for a general multi band spectrum sensing scenario. We mention [11] and [26] here as possible interesting directions for future studies where the recency-based policies proposed in this paper could be applied assuming either side information or different optimization goals. In [11] the authors formulate the learning problem of joint sensing and access as a RMAB problem with side observations and assume that the SUs sensing performance (detection probability and false alarm probability) is known. The paper then proposes using an UCB-type policy for solving the problem. Another interesting application of RMAB formulation in CR was given in [26], where the authors propose a PAC learning algorithm that determines the amount of exploration needed in RMAB in order to balance between the energy consumption of sensing plus probing and access.

## III. Problem Formulation

### A. System model

At time instant $n$ the CR network senses (and possibly accesses) a frequency band $k$, $k = 1, ..., K$, and observes an achieved data rate (reward) $x_k(n)$ with an unknown mean $\mu_k$. In this paper the data rates are assumed to be either i.i.d. in time or evolve as a stationary 2-state Markov process. The rates $x_k(n)$ are assumed to be $[0, 1]$ bounded which can be achieved by normalizing the true data rates (bits/s) with the highest Shannon capacity among the bands. Also, it is assumed that the SUs have a way of estimating and feeding back the achieved data rates to a central node, e.g. a fusion center that runs and maintains the sensing policy for the whole network or for a small subnetwork. All bands are assumed to evolve independently from each other.

### B. Objective

The objective of this paper is to develop a simple sensing policy for the CR that achieves an order optimal tradeoff between exploration and exploitation. Quantitatively, the success of a policy $\pi$ can be measured by its expected weak regret $\mathrm{E}\big[R^\pi(n)\big]$. The weak regret of a policy $\pi$ is defined as the difference of the total payoff achieved by the policy $\pi$ and the total payoff achievable by the optimal single arm policy $\pi^*$. Mathematically this can be expressed as

$$\mathrm{E}\big[R^\pi(n)\big] = \sum_{k:\mu_k \neq \mu_{k^*}} \Delta_k E[M_k^\pi(n)] \quad (1)$$

where $n$ is the time index, $\Delta_k = \mu_{k^*} - \mu_k$ is the optimality gap of band $k$, $M_k^\pi(n)$ is the number of times band $k$ has been sensed up to time $n$ using policy $\pi$ and $k^* = \arg\max_k \mu_k$. In order to simplify the notation the superscript $\pi$ will be dropped for the rest of the paper.

### C. Discussion on the System Model

In practice the notion of the single best frequency band may be ambiguous. Since the SUs may be scattered in space they experience different channel fading and consequently obtain different data rates in different locations. For the same reason, the probabilities of detection and false alarm may be different at different locations. Taking these factors into account the optimal sensing policy becomes a function of the access policy (access policy tells who will get access to the possibly idle band) and the employed sensing scheme. Such joint optimization of sensing and access in CR has been considered for example in [11], [27], [28]. Also, in practice the rewards from different bands are not necessary independent since high power primary transmissions (such as TV-transmission) may cause out-of-band interference to the neighboring bands.

In many potential CR settings, the data rates (rewards) are non-stationary. For example the obtained throughputs depend on the amount of traffic in the primary network that may vary between peak and off-peak hours. Also the time-frequency-location varying nature of the wireless channels and user mobility will in practice cause the secondary users data rates to be non-stationary. In these situations a sensing policy assuming stationarity has to be occasionally restarted afresh. Alternatively, in exploration bonus based policies the exploration bonus could be tuned so that after a fixed period exploration becomes more attractive again. In this paper, however, we concentrate on the stationary problem alone.

## IV. THE PROPOSED POLICY

### A. Policy for i.i.d. rewards

In this section we propose a spectrum sensing policy for CR when the data rates (rewards) are $[0,1]$ bounded and i.i.d. in time. The proposed sensing policy is an index policy that contains an exploitation promoting sample mean term and an exploration promoting bonus. Exploration bonuses are awarded to the band according to recency, such that bands that have not been sensed for a long time will get higher bonus and consequently will be more likely to be sensed in the near future. The proposed sensing policy is detailed in Fig. 1 and equation (2).

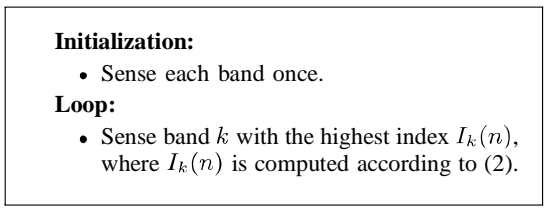

**Initialization:**
- Sense each band once.

**Loop:**
- Sense band $k$ with the highest index $I_k(n)$, where $I_k(n)$ is computed according to (2).

Fig. 1. The proposed sensing policy for i.i.d. rewards.

The index of band $k$ at time $n$ is given as

$$I_k(n) = \bar{x}_k(n) + g\left(\frac{n}{\tau_k(n)}\right), \quad (2)$$

where $\bar{x}_k(n)$ is the sample mean of the rewards from band $k$, $g(n/\tau_k(n))$ is the exploration bonus and $\tau_k(n)$ is the last time instant when band $k$ was sensed. The sample mean is computed as $\bar{x}_k(n) = \frac{1}{M_k(n)} \sum_{m=1}^{M_k(n)} x_k(u_m^{k,n})$, where $u_m^{k,n}$ is the sequence of length $M_k(n)$ of the sensing time instants of band $k$ up to time $n$.

The exploration bonus $g(x), x > 0$ is a concave, strictly increasing and unbounded function such that $g(1) = 0$. An example of such exploration bonus would be $g(x) = \sqrt{\ln(x)}$. Fig. 2 shows the exploration bonus of a suboptimal band as a function of time when using it in the proposed sensing policy. It can be seen that the zeroes of the exploration bonus indicate the sensing time instants of the subband and that the time instants when the subband is sensed tend to grow exponentially as sensing focuses on the band giving the highest rewards.

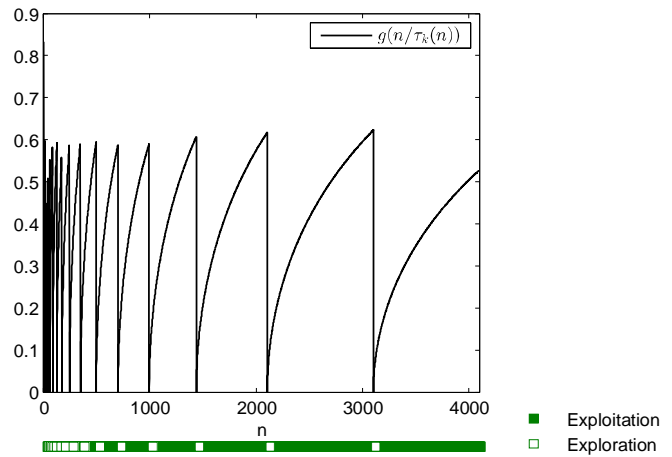

Fig. 2. The exploration bonus of a *suboptimal* channel as a function of time in a two frequency band scenario for $g(n/\tau_k(n)) = \sqrt{\ln(n/\tau_k(n))}$. In this example the difference between the expected rewards of the optimal band and the suboptimal band is $\Delta = \mu^* - \mu = 0.65$. The time interval between two consecutive sensing time instants (the zeroes) of the suboptimal band tends to grow exponentially in time. In other words it means that the lengths of the exploitation epochs grow exponentially. Since the time intervals between two consecutive sensings of a suboptimal channel grow exponentially it means that the total number of sensings must grow logarithmically in time.

The effect of the choice of the exploration bonus on the weak regret becomes now intuitive. Employing a $g(x)$ that increases fast from 0 means that asymptotically all bands (including the suboptimal ones) will be sensed more often compared to a choice of $g(x)$ that increases slowly. Fast growing $g(x)$ will lead to aggressive exploration whereas slow growing $g(x)$ leads to aggressive exploitation. With aggressive exploration the policy's asymptotic weak regret can be reached fast but its value will be high, whereas with aggressive exploitation the convergence to the asymptotic regret will be slow but the regret itself will be small. This trade-off is dealt in more detail in section V.

The asymptotic weak regret of the policy in Fig. 1 is summarized in Theorem 1.

**Theorem 1.** *The asymptotic weak regret of the policy in Fig. 1 when the rewards are i.i.d. is*

$$\mathrm{E}[R(n)] = \sum_{k:\mu_k \neq \mu_{k^*}} \Delta_k O\left(\ln(n)\right). \quad (3)$$

*Proof:* See appendix A ∎

### B. Markovian rewards

In practice the assumption that the state of a frequency band evolves independently in time may not be always valid. Following the line of [11], [24], [29], [30] we model the evolution of the state of the spectrum band with a 2-state Markov chain (see Fig. 3), also known as the Gilber-Elliot model. We propose a small modification for the sensing policy in Fig. 1 so that provably asymptotically logarithmic weak regret will be attained. The condition for the logarithmic weak regret is that the Markov chain of the Gilbert-Elliot model is ergodic (irreducible, aperiodic and such that all states are positive recurrent). We employ the Gilbert-Elliot model due to its correspondence with dynamic spectrum access scenario. In a real CR network the SUs would be equipped with signal detectors which perform binary hypothesis testing on the availability of a particular frequency band. Assuming that the detectors are well designed the secondary users may then access the band only when the state of the spectrum is detected to be idle. The two states in the Gilbert-Elliot model correspond to these two possible outputs from the detector: state 1 indicating that the band is occupied and state 0 indicating that the band is idle. Note, however, that the actual observed rewards (data rates) from these two states can be any two values between 0 and 1.

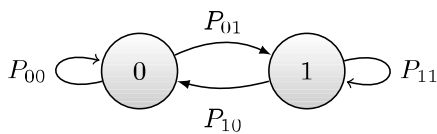

Fig. 3. Markov chain used to model the temporal dependency of the state of a band. In this paper state 1 denotes that the band is occupied and state 0 that the band is idle.

Our policy construction for Markovian rewards is inspired by [9] and [11] by making use of the regenerative property of Markov chains. In particular, we constrain the periods of exploration and exploitation of each frequency band to be an integer multiple of a full *regenerative cycle*. Starting from state $S$ a regenerative cycle of a Markov chain is a sequence of states before the chain returns back to state $S$. See Fig. 5 for an illustration of regenerative cycles. When the chain is irreducible and aperiodic the lengths of the regenerative cycles of a given state are i.i.d. (see e.g. [31],[32]). This "trick" of breaking the observed states of a Markov chain into regenerative cycles is often used in order to employ the theory of independent random variables, for example, when proving the strong law of large numbers for Markov chains [32]. The proposed policy for Markovian rewards is shown in Fig. 4. Fig. 5 illustrates the regenerative cycles during the first 12 sensing time instants in a 2 band scenario with the proposed sensing policy.

The main difference to the policy in Fig. 1 is that each time a band is selected for sensing it will be sensed until at least one full regenerative cycle is observed. This also helps in making the analysis of the weak regret in appendix B simpler and intuitive.

Since the Bernoulli distribution is a special case of two-state Markov chain, the proposed sensing policy in Fig. 4 can naturally also achieve asymptotically logarithmic weak regret with Bernoulli distributed rewards. However, in practice, due to the fact that the policy in Fig. 4 forces the CR to sense the same band for at least one full regenerative cycle, it often tends to perform slightly worse with i.i.d. rewards than the policy in Fig. 1.

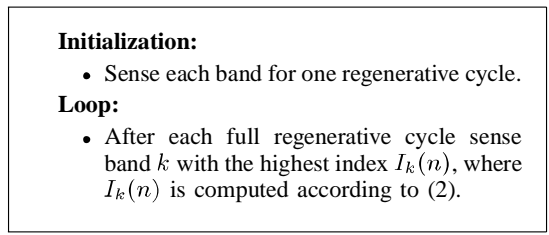

Fig. 4. The proposed sensing policy for rewards evolving according to a 2-state Markov chain.

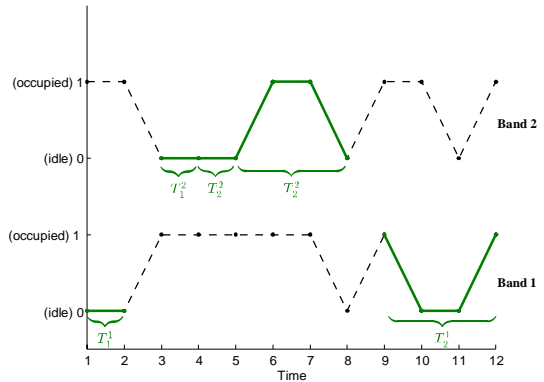

Fig. 5. Illustration of the first 12 sensing time instants using the proposed sensing policy. The number of bands is 2 and they are assumed to follow a 2-state Markov-chain. The horizontal axis indicates time and the vertical axis indicates the states of the two bands. The dashed line indicates the state evolution of the band. The green solid line shows the sensed states by the SU. The lengths of the regenerative cycles of band 1 are denoted as $T_1^1, T_2^1, \ldots$ and similarly for band 2. At time instants 1-2 the SU senses one regenerative cycle at band 1. During time instants 3-8 the SU senses 3 regenerative cycles at band 2 and at time instants 9-12 the SU senses band 1. The indices of the bands are recalculated after each observed full regenerative cycle. Consequently the decision to keep sensing the same band or to switch to another is also made after each observed regenerative cycle. In this example these index update and decision time instants correspond to 2,4,5,8 and 12.

The following theorem summarizes the asymptotic weak regret of the policy in Fig. 4.

**Theorem 2.** *The asymptotic weak regret of the policy in Fig. 4 when the rewards are evolving according to a 2-state Markov process is logarithmic, i.e.,*

$$\mathrm{E}[R(n)] = \sum_{k:\mu_k \neq \mu_{k^*}} \Delta_k O(\ln(n)). \qquad (4)$$

*Proof:* See appendix B ∎

### C. Intuition on asymptotic logarithmic weak regret

Next we provide the intuition why the proposed policies in Fig. 1 and Fig. 4 attain asymptotically logarithmic weak regret. Detailed proofs can be found in appendices A and B.

As was seen in (1) the expected weak regret depends on the number of times that a suboptimal frequency band is sensed. In order to show that the weak regret is logarithmic one needs to show that the expected number of sensings of any suboptimal band is upper bounded logarithmically. Our analysis is based on investigating the interval between two consecutive sensing time instants of any suboptimal band (the instants of the zeroes seen in Fig. 2) after sufficient number of samples from each band has been accumulated. It is shown that the time difference between the first and the $j^{\text{th}}$ (for large enough $j$) sensing time instant of a suboptimal band grows exponentially in time and that this fact consequently leads to logarithmically growing weak regret.

Since the rewards are bounded and since by definition $g(1) = 0$ and $g(x) \geq 0$, at any given time for any band there always exists a future time instant when the band will be sensed again. In other words the policy never completely stops exploring and consequently each band will be asymptotically sensed infinitely often. After sufficient amount of exploration due to the strong law of large numbers the indices of a suboptimal band and the optimal band may be approximated respectively as

$$\lim_{n \to \infty} I_k(n) = \mu_k + g(n/z_{k,j})$$
$$\lim_{n \to \infty} I_{k^*}(n) = \mu_{k^*} + g(n/\tau_{k^*}(n)) \geq \mu_{k^*},$$

where $z_{k,j}$ is the time instant of the $(j)^{\text{th}}$ sensing of (suboptimal) band $k$. The latter inequality follows from the fact that the exploration bonus $g(n/\tau_{k^*}(n))$ is always greater or equal to zero. This is to say that the sample mean terms behave asymptotically like constants and that for sufficiently large $n$ the times of exploration and exploitation are practically controlled by the exploration bonus $g(n/\tau_k(n))$. Hence, asymptotically a suboptimal band $k$ will not be sensed sooner than when $I_k(n) \geq \mu_{k^*}$, i.e., when

$$\mu_k + g(z_{k,j+1}/z_{k,j}) \geq \mu_{k^*} \;\Rightarrow\; z_{k,j+1} \geq z_{k,j} g^{-1}(\Delta_k).$$

Since $g(1) = 0$ and $g(x)$ strictly increasing, concave and unbounded on $x \in [1, \infty)$, the inverse function $g^{-1}(0) = 1$ and $g^{-1}(y)$ is strictly increasing, convex and unbounded on $y \in [0, \infty)$. Consequently $g^{-1}(\Delta_k) \geq 1$, where the equality holds only when $\Delta_k = 0$. For the $(j+2)^{\text{th}}$ sensing $z_{k,j+2} \geq z_{k,j}(g^{-1}(\Delta_k))^2$ and so on for the $(j+M)^{\text{th}}$ sensing time instant $n \triangleq z_{k,j+M} \geq z_{k,j}(g^{-1}(\Delta_k))^M$. This implies that asymptotically

$$M \leq \frac{\ln(n) - \ln(z_{k,j})}{\ln(g^{-1}(\Delta_k))}. \tag{5}$$

The above intuition is based on the strong law of large numbers which states that there exists a finite time instant when approximating the sample mean terms with the true mean values is good enough and after which the number of suboptimal sensings will increase only logarithmically. This is illustrated graphically in Fig. 6, where after time instant $n_0$ the number of sensings of a suboptimal band grows logarithmically.

While much of the recent work in this area has concentrated on finding finite time upper bounds on the expected weak regret, the bounds derived in this paper are asymptotic by nature. Most of the recent derivations of finite time weak regret bounds in the literature owe themselves to the framework first given in [16] by using Chernoff/Hoeffding type concentration inequalities. The policies proposed in this paper are not applicable to this proofing technique, since the exploration bonuses here are based on recency - not confidence explicitly. However, in Appendices A and B, we show that for every run of the proposed policy with probability 1, there exists a finite time instant after which the weak regret grows logarithmically. In practice, this means that whenever the policy is set running, the weak regret converges with probability 1 to a logarithmic rate in a finite time.

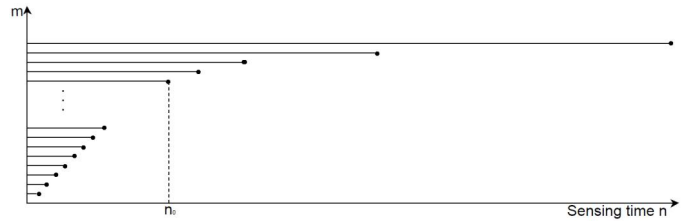

Fig. 6. Example of the asymptotic behavior of the number of suboptimal sensings for one sample path. The x-axis indicates the suboptimal sensing time instants and the y-axis indicates the ordinal of the sensing (the number of suboptimal sensings). The length of each horizontal line indicates the sensing time instant for that the $m$th sensing. For the first sensings the number of suboptimal sensings might in the worst case grow linearly with time, but with probability one for large enough $n$ (here $n_0$) the growth will slow down to logarithmic.

## V. CONSTRUCTING THE EXPLORATION BONUS

In this section we present how to find the exploration bonus $g(n/\tau_k(n))$. The goal is to find a feasible exploration bonus that brings the leading constant of the asymptotic weak regret in the i.i.d. case as close as possible to the asymptotic lower bound given in [12]. Here by the term "feasible", we mean an exploration bonus within the proposed class of recency-based policies that is also simple to compute. Deriving a policy that would match the lower bound of Lai and Robbins in [12] is out of the scope of this paper, since to that end the proposed policy would need nontrivial estimates of the Kullback-Leibler divergences between the reward distributions. Here we will only consider the case of (general) independent rewards and Bernoulli rewards, since the lower bound of [12] applies to those cases. Interestingly, the lower bound for the weak regret in the general restless Markov case is not known, but deriving is omitted in this paper.

### A. Independent rewards: general case

In [12], Lai and Robbins showed that for any consistent policy $\pi$, the following lower bound holds for the number of sensings of a suboptimal band:

$$\liminf_{n \to \infty} \frac{\mathrm{E}[M^\pi(n)]}{\ln(n)} \geq \sum_{k:\mu_k < \mu^*} \frac{1}{d(\theta_k || \theta^*)}, \tag{6}$$

where $d(\theta_k || \theta^*)$ is the Kullback-Leibler divergence of the reward distribution $\theta_k$ of the suboptimal band $k$ to the reward distribution $\theta^*$ of the optimal band and $M^\pi(n)$ is the total number of suboptimal sensings by policy $\pi$. We would like the asymptotic leading constant $1/\ln(g^{-1}(\Delta_k))$ of the weak regret of our policy in (5) to be as close as possible (from above) to $1/d(\theta_k || \theta^*)$. Hence, we are looking for a $g^{-1}(\Delta_k)$

that satisfies

$$g^{-1}(\Delta_k) \leq \exp(d(\theta_k||\theta^*)), \quad (7)$$

as close as possible to equality. Since $\exp(d(\theta_k||\theta^*))$ is a convex function in the pair $(\theta_k, \theta^*)$ [33], one can conclude that it is best approximated from below by $g^{-1}(\Delta_k)$ that is convex. Consequently, since $g^{-1}(\Delta_k)$ needs to be convex, the exploration bonus $g(n/\tau_k)$ itself should be concave. This is also supported by intuition and by keeping in mind Fig.2: If the difference in the mean reward between the optimal band and the suboptimal band is small (close to 0), one can afford to increase the exploration bonus fast in the beginning since both the optimal and suboptimal bands need to be explored many times in order to find out which one of them has a larger expected reward. On the other hand, if the difference in the mean reward is big (close to 1), one can still increase the exploration bonus fast in the beginning but should gradually slow it down in order to inhibit excessive exploration.

For deriving an exploration bonus for bounded i.i.d. rewards we observe the following result due to Pinsker's inequality:

**Theorem 3.** *Let $X, Y \in [0, 1]$ be continuous random variables with integrable probability density functions $p_X(\lambda)$ and $p_Y(\lambda)$. Let $\Delta = |\mathrm{E}[X] - \mathrm{E}[Y]|$. Then*

$$\frac{1}{2}\Delta^2 \leq d(p_X(\lambda)||p_Y(\lambda)), \quad (8)$$

*where $d(\cdot||\cdot)$ denotes the Kullback-Leibler divergence.*

*Proof:*

$$\begin{aligned}\Delta =& \Big|\int_0^1 \lambda(p_X(\lambda)-p_Y(\lambda))d\lambda\Big| \leq \int_0^1 \lambda\Big|(p_X(\lambda)-p_Y(\lambda))\Big|d\lambda \\ \leq& \int_0^1 \Big|(p_X(\lambda)-p_Y(\lambda))\Big|d\lambda = 2\delta(X,Y) \\ \leq& \sqrt{2d(p_X(\lambda)||p_Y(\lambda))},\end{aligned} \quad (9)$$

where $\delta(X, Y)$ is the total variation distance between $X$ and $Y$. Equation (8) then follows from (9). The second inequality is due to the fact that we are integrating over the unit interval $0 \leq \lambda \leq 1$. The last step in (9) is due to Pinsker's inequality (see e.g. [34]). With practically the same arguments (and by replacing integrals with sums), the above result can be derived for $X$ and $Y$ that are $[0, 1]$ bounded discrete random variables. ∎

Using the fact that $\frac{1}{2}\Delta_k^2 \leq d(\theta_k||\theta^*)$ from Theorem 3, we can find the exploration bonus for the case when the rewards are assumed to be independent and bounded in $[0, 1]$. Keeping in mind that asymptotically at the sensing time instants $n/\tau_k(n) = g^{-1}(\Delta_k)$ and requiring that $\ln(g^{-1}(\Delta_k)) = \frac{1}{2}\Delta_k^2$, it is possible to find $g(n/\tau_k(n))$ to be

$$g\left(\frac{n}{\tau_k(n)}\right) = \sqrt{2\ln\left(\frac{n}{\tau_k(n)}\right)}. \quad (10)$$

Using (10) as the exploration bonus in the proposed policy (listed in Fig. 1), the leading constant of the asymptotically logarithmic weak regret will approach $\frac{2}{\Delta_k^2}$ according to (5).

For illustration purposes we simulate the proposed policy with the exploration bonus given in (10) in a two frequency band scenario. The rewards (data rates) from band 1 are i.i.d. uniform between $[0, \frac{1}{2}]$ and the rewards from band 2 are i.i.d. uniform between $[\frac{1}{2}, 1]$. With these the expected rewards are $\mu_1 = \frac{1}{4}$ and $\mu_2 = \frac{3}{4}$ so that $\Delta_1 = \frac{1}{2}$. Fig. 7 plots the rate of change of the number of sensings of band 1 with respect to $\ln(n)$. As expected by (5), the rate of change asymptotically converges to $\frac{2}{\Delta_1^2} = 8$.

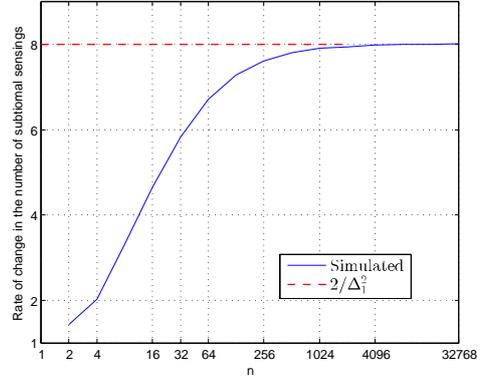

Fig. 7. The simulated rate of change of the number of sensings of band 1 with respect to $\ln(n)$. The curve has been averaged over 1000 Monte Carlo simulations. The simulated rate of change approaches the theoretical value $2/\Delta_1^2 = 8$.

### B. Bernoulli rewards

Next we construct the exploration bonus for the case when the rewards are known to be independent Bernoulli distributed. By assuming the data rates to be Bernoulli variables (in addition to being independent), it is possible to obtain a tighter version of the Pinsker's inequality and hence use greedier exploitation than what would be achieved by the exploration bonus in (10).

**Theorem 4.** *Assume two Bernoulli random variables $X$ and $Y$ in $\{0, 1\}$. Let the success probability (probability of value 1) of $X$ be $p_x$ and the success probability of $Y$ be $p_y$. Furthermore, let $\Delta = |\mathrm{E}[X] - \mathrm{E}[Y]| = |p_x - p_y|$. Then*

$$2\Delta^2 \leq d(p_x||p_y) \quad (11)$$

*Proof:*

$$\begin{aligned}\Delta =& |(p_x - p_y)| = \frac{1}{2}\big(|p_x - p_y| + |(1-p_x) - (1-p_y)|\big) \\ =& \delta(X,Y) \leq \sqrt{\frac{1}{2}d(p_x||p_y)},\end{aligned} \quad (12)$$

where $\delta(X, Y)$ is the total variation distance between $X$ and $Y$ and the last inequality is due to Pinsker's inequality. Hence solving for $d(p_x, p_y)$ yields equation (11). ∎

Note that the theorem above holds for any Bernoulli process (and not only when the rewards are either 0 or 1). Employing Theorem 4 we obtain the tailored exploration bonus for Bernoulli distributed rewards as

$$g\left(\frac{n}{\tau_k(n)}\right) = \sqrt{\frac{1}{2}\ln\left(\frac{n}{\tau_k(n)}\right)}. \quad (13)$$

Theorem 3 provides a lower bound for the KL-divergence of all independent rewards bounded in $[0, 1]$, whereas Theorem 4 provides a tighter lower bound when the rewards are also known to be Bernoulli. Hence, when there is a priori information about the type of the reward distributions (in this case, Bernoulli) it is possible to use an exploration bonus that favors more aggressive exploitation. In other words, when the reward distribution is known apart from its expectation, it is possible to more carefully balance the trade-off between the convergence to the optimal frequency band and the achieved asymptotic regret.

## VI. SIMULATION EXAMPLES

In this section we illustrate the performance of the proposed sensing policies in various scenarios with independent and Markovian data rates (rewards) by simulations. Here the performance measure is the number of suboptimal sensing, i.e., the number of times when other than the band with the highest stationary expectation was sensed. The performance of the policies is compared against 3 other cutting edge policies KL-UCB [17], DSEE [10] and RCA [9]. In order to guarantee the DSEE and RCA policies to achieve finite time logarithmic weak regret, one needs to define appropriate parameter values for them (parameter $D$ for the DSEE and parameter $L$ for the RCA). To this end one would need to know certain non-trivial upper bounds for the parameters of the underlying Markov processes. This information may not be available in practice. In addition, it has been empirically shown in [10], [9] that these theoretical parameter values, although being sufficient, are not necessary and that often better performance is achieved with lower values of $D$ and $L$. However, setting $D$ and $L$ to low fixed values might improve their performance in some scenarios but might lead to significant performance degradation in other scenarios. According to [10], by letting the policy parameter $D$ of the DSEE to slowly grow with time eliminates the need for any a priori system knowledge with an arbitrary small sacrifice in the asymptotic weak regret. In the simulations we have set the DSEE policy parameter $D(n) = \ln(n)$ with the goal of obtaining the best possible overall performance in different scenarios. In our experiments, it provided the most stable outcome with a good finite performance. In all the scenarios considered here, the weak regret with $D = \ln(n)$ was practically the same as with $D = 10$ that was used in the simulations of [10]. Similarly according to [9], letting the parameter $L$ of the RCA policy to grow slowly in time will not sacrifice the asymptotic regret much. In the simulations of RCA we have used $L(n) = \ln(n)$ and $L = 1$, where the value $L = 1$ is taken from the simulations of [9]. All the simulations are $2^{15} (\approx 32000)$ sensing time instants long and the presented curves are averages of 10000 independent runs. The curves have been normalized by $1/\ln(n)$ in order to illustrate convergence to a logarithmic rate.

### A. Independent rewards

First we simulate the performance of the proposed sensing policy when the rewards are independent. We consider the case when the rewards are Bernoulli distributed and the case when the rewards can obtain any values between [0,1].

Fig. 8 shows the expected number of suboptimal sensing in a 5 band scenario. The availability of the band is assumed to be Bernoulli distributed such that the reward is 1 when the band is sensed idle and accessed and 0 when the band is sensed occupied. It is assumed that trying to access an occupied band would cause a collision between the SU and the PU and produce no throughput. The average data rates of the bands are then $\mu = [0.1, 0.7, 0.5, 0.6, 0.8]$. In the implementation of the KL-UCB we have assumed Bernoulli distribution, hence it represents here an asymptotically optimal policy. For the proposed policy we show the results using the exploration term in (13) for Bernoulli rewards. It can be seen that in this case the proposed sensing policy performs close to the KL-UCB policy while being computationally much simpler. The DSEE policy has a significant drop in its performance around after 10000 sensing time instants. This seems to be due to the deterministic nature of the exploration epochs, which in this scenario often occurs around $n \approx 10000$. In this scenario the RCA policy takes the longest time before converging to a logarithmic rate.

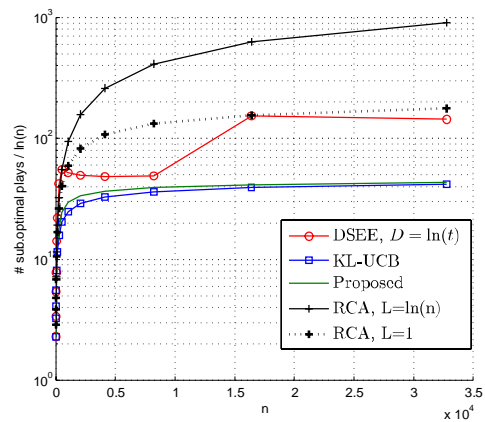

Fig. 8. Mean number of suboptimal sensings with Bernoulli rewards. The expected rewards of the bands are $\mu = [0.1, 0.7, 0.5, 0.6, 0.8]$. It can be seen that in this case the proposed sensing policy gives almost as good performance as the KL-UCB policy that has been shown to be asymptotically optimal. However, the proposed policy is much simpler than the KL-UCB.

Fig. 9 shows the average number of suboptimal sensing in a 2-band scenario where the reward distributions of the bands are shown in Fig. 10. In this case, the KL-UCB policy for Bernoulli rewards is no longer asymptotically optimal. However, according to the authors of [17] it should still achieve good performance with general [0,1] bounded rewards. For the DSEE we have again used the parameter value $D = \ln(n)$ and for the RCA the parameter $L = \ln(n)$ and $L = 1$. In order to simulate the RCA policy in this scenario we have assumed that the SU would be capable of distinguishing between the 101 possible rewards (states) shown on the horizontal axis of Fig. 10. For the proposed, policy we have used the exploration bonus given in (10) optimized for general i.i.d. [0,1] bounded rewards and the exploration bonus optimized for Bernoulli rewards. We also observed that the proposed policy opti-mized for Bernoulli rewards achieves excellent performance even when the reward distributions are not Bernoulli. In this scenario the proposed policy assuming Bernoulli rewards and the DSEE policy achieve the lowest number of suboptimal sensings.

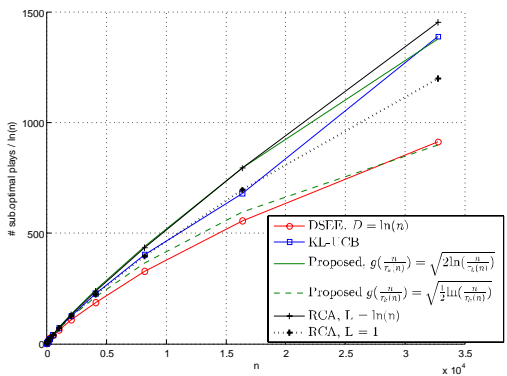

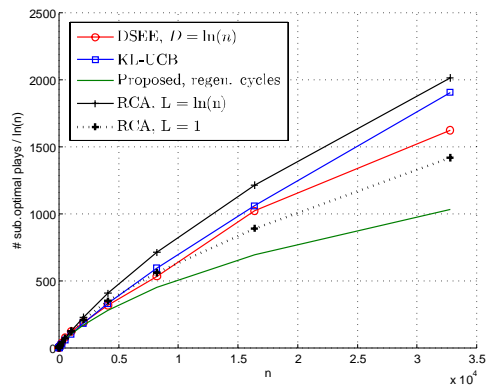

Fig. 9. Average number of suboptimal sensings with IID rewards with reward PMFs shown in Fig. 10. The KL-UCB policy is the one optimized for Bernoulli rewards. For the proposed sensing policy we have used the exploration bonus given in (10) and (13). It can be seen that the exploration bonus optimized for Bernoulli rewards gives still excellent performance even though the actual rewards are not Bernoulli distributed.

Fig. 11. The simulated mean number of suboptimal sensings in a scenario with very slowly varying spectrum with 10 bands and Markovian rewards. The transition probabilities are $P_{10} = [0.01, 0.01, 0.02, 0.02, 0.03, 0.03, 0.04, 0.04, 0.05, 0.05]$ and $P_{01} = [0.08, 0.07, 0.08, 0.07, 0.08, 0.07, 0.02, 0.01, 0.02, 0.01]$ making the typical state evolutions of the bands to be ...0,0,0,0,...0,1,1,1,1,.... Due to the slowly varying states of the spectrum this scenario would be highly attractive for CR. The KL-UCB policy has is the one optimized for Bernoulli rewards and in the DSEE policy we have set the parameter $D = \ln(t)$. For the RCA we have used $L = \ln(n)$ and $L = 1$. In the proposed policy the exploration bonus is $g\left(\frac{n}{\tau_k(n)}\right) = \sqrt{2\ln\left(\frac{n}{\tau_k(n)}\right)}$. In this scenario the proposed policy achieves clearly the lowest number of sensings on suboptimal bands.

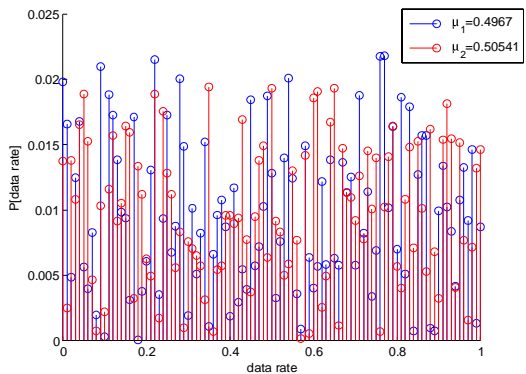

Fig. 10. The data rate (reward) PMFs of the two bands simulated in Fig. 9. The expected rewards are $\mu_1 = 0.4967$ and $\mu_2 = 0.5054$. This corresponds to a difficult scenario where the two bands have almost the same expected rewards.

*B. Markovian rewards*

Next we present the simulation results with Markovian rewards of the proposed sensing policy listed in Fig. 4. Fig. 11 shows the expected number of suboptimal sensings when there are 10 bands whose availability for secondary use evolves according to a 2-state Markov chain (i.e. the Gilbert-Elliot model). Also in this scenario when the band is in state 1 the band is occupied by the primary user and when the band is in state 0 the band is idle. The reward from sensing a band that is idle is 1 whereas sensing an occupied band produces 0 reward. The transition probabilities of the bands are $P_{10} = [0.01, 0.01, 0.02, 0.02, 0.03, 0.03, 0.04, 0.04, 0.05, 0.05]$ and $P_{01} = [0.08, 0.07, 0.08, 0.07, 0.08, 0.07, 0.02, 0.01, 0.02, 0.01]$. The corresponding stationary expected rewards are $\mu = [0.83, 0.11, 0.80, 0.30, 0.67, 0.20, 0.71, 0.22, 0.13, 0.27]$. This scenario corresponds to a case where the state of the spectrum evolves slowly between occupied and unoccupied, i.e., the spectrum is either persistently idle or persistently occupied. Such scenario would be very attractive for opportunistic spectrum use. In the proposed, policy we have used the exploration bonus $g\left(\frac{n}{\tau_k(n)}\right) = \sqrt{2\ln\left(\frac{n}{\tau_k(n)}\right)}$. Fig. 11 shows that the proposed policy achieves uniformly lowest number of suboptimal sensings compared to the other three policies.

Fig. 12 shows the average number of suboptimal sensings for the proposed sensing policy in a scenario where the state of the spectrum is highly dynamic. In this scenario the state transition probabilities are close to 1 causing the bands to all the time alternate between idle and occupied state, which makes this scenario less attractive for practical cognitive radio employment. In the DSEE we have again set $D = \ln(n)$ and in the RCA $L = \ln(n)$ and $L = 1$. In the proposed policy we have used the exploration bonus $g\left(\frac{n}{\tau_k(n)}\right) = \sqrt{2\ln\left(\frac{n}{\tau_k(n)}\right)}$. In this scenario the proposed policy is on par with the RCA with $L = 1$ while the DSEE has the lowest number of suboptimal sensings.

VII. CONCLUSIONS

In this paper we have proposed asymptotically order optimal sensing policies for cognitive radio that carry out recency-based exploration through the use of carefully developed exploration bonuses. We have proposed policies for the cases in which the state of the spectrum evolves independently from the past and when the state evolves as a 2-state Markov process. The proposed policies are built upon the idea of recency-based exploration bonuses that force each band to be sensed infinitely many times while ultimately pushing the exploration instants of suboptimal bands exponentially far apart from each other. We have proved using analytical tools that the proposed policies attain asymptotically logarithmic weak regret when the bounded rewards are independent and when they are Markovian. Furthermore, we have shown that when there is information about the type of the secondary user throughput distributions, it is possible to construct policies with better performance. Our simulation results have shown that the proposed policies provide typically performance gains over the state-of-the-art policies. The simulation results have also indicated that the expected weak regret would also be uniformly logarithmic.

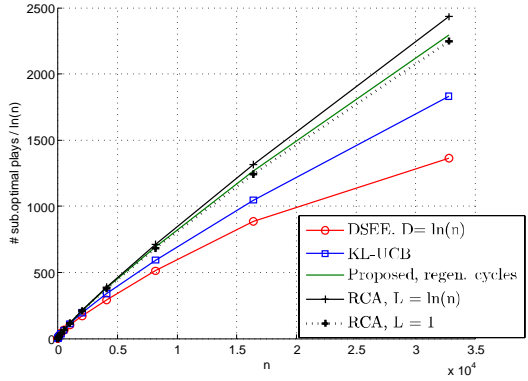

Fig. 12. The simulated mean number of suboptimal sensings in a highly dynamic spectrum with 5 bands and Markovian rewards. The transition probabilities are $P_{10} = [0.95, 0.97, 0.94, 0.91, 0.96]$ and $P_{01} = [0.94, 0.93, 0.91, 0.97, 0.91]$ making the typical state evolutions of the bands to be ...,0,1,0,1,0,.... Since the spectrum is highly dynamic this scenario would be less attractive for cognitive radio employment. The KL-UCB policy has is the one optimized for Bernoulli rewards and in the DSEE policy we have set the parameter $D$ to $D = \ln(t)$. For the RCA we have used $L = \ln(n)$ and $L = 1$. In the proposed policy we have used the exploration bonus $g\left(\frac{n}{\tau_k(n)}\right) = \sqrt{2\ln\left(\frac{n}{\tau_k(n)}\right)}$. In this scenario the proposed policy and the RCA perform equally well while the DSEE has the best performance.

## APPENDIX A
## PROOF OF THEOREM 1

In this section, we give a formal proof that the proposed policy in Fig. 1 attains asymptotically logarithmic weak regret when the rewards are independent. The proof is based on the fact that the event $\{\bar{x}(n) > \bar{x}^*(n)\}$ happens only a finite number of times when $n \to \infty$ and henceforth asymptotically the suboptimal band will be sensed only when its exploration bonus has become large enough (see Fig. 2). Asymptotically, this event happens at an exponentially decreasing rate with probability 1. In order to keep the notation simple the derivation is given for the case of two bands, however, without loss of generality. The result will generalize to multiple frequency bands by comparing each suboptimal band against the optimal band separately. Since the optimal band is asymptotically sensed exponentially more often than any of the suboptimal bands, the asymptotic weak regret will be logarithmic.

Here we use the following notation for the suboptimal band: $\bar{x}(n) = \frac{1}{M(n)}\sum_{i=1}^{M(n)} x(u_i)$, where $\{u_i\}$ is the sequence of length $M(n)$ of the sampling instants up to time $n$ and $\mu = \mathrm{E}[x(n)]$. Variables with $^*$ denote the corresponding values for the optimal band. Furthermore, $\Delta = \mu^* - \mu > 0$ is the difference of the true mean of the optimal band and the true mean of the suboptimal band.

Next we show that for each sample path (run of the sensing policy) there exists with probability 1 a time instant $n_0 < \infty$ when the suboptimal band is sensed and after which the event $\{\bar{x}(n) \geq \bar{x}^*(n)\ \forall n > n_0\}$ does not take place any more. In other words, one can show that for any $\epsilon > 0$ and large enough $n$ the sample mean of the optimal band will be larger by $(1-\epsilon)\Delta$ than the sample mean of the suboptimal band. After this point the explorations of the suboptimal band will be almost surely dictated by the exploration bonus. To this end we use the following lemma by Kolmogorov (see e.g. [35] p. 27):

**Lemma 1.** *(Kolmogorov's strong law).* Let $X_1, X_2, \ldots$ be independent with means $\mu_1, \mu_2, \ldots$ and variances $\sigma_1^2, \sigma_2^2, \ldots$. If the series $\sum_{i=1}^{\infty} \sigma_i^2/i^2$ converges, then

$$\frac{1}{n}\sum_{i=1}^{n}(X_i - \mu_i) \overset{\mathrm{a.e.}}{\to} 0, \qquad (14)$$

*where a.e. stands for almost everywhere.*

*Proof:* See [36] p. 590. ∎

Note that Lemma 1 implies that for any $\epsilon > 0$

$$\mathrm{P}\left\{\left|\frac{1}{n}\sum_{i=1}^{n}(X_i - \mu_i)\right| > \epsilon, \mathrm{i.o.}\right\} = 0, \qquad (15)$$

where i.o. stands for infinitely often. Now, for any $\epsilon > 0$

$$\mathrm{P}\left\{\bar{x}^*(n) - \bar{x}(n) \leq (1-\epsilon)\Delta,\ \mathrm{i.o.}\right\}$$
$$= \mathrm{P}\left\{\bar{x}^*(n) - \bar{x}(n) - (\mu^* - \mu) \leq -\epsilon\Delta,\ \mathrm{i.o.}\right\}$$
$$= \mathrm{P}\left\{\frac{1}{m^*}\sum_{k=1}^{m^*}(x^*(u_k^*) - \mu^*) + \frac{1}{m}\sum_{i=1}^{m}(\mu - x(u_i)) \leq -\epsilon\Delta,\ \mathrm{i.o.}\right\}$$
$$= \mathrm{P}\left\{\frac{1}{m^*}\sum_{k=1}^{m^*}(x^*(u_k^*) - \mu^*) + \frac{\epsilon\Delta}{2}\right.$$
$$\left. + \frac{1}{m}\sum_{i=1}^{m}(\mu - x(u_i)) + \frac{\epsilon\Delta}{2} \leq 0,\ \mathrm{i.o.}\right\},$$

where $u_k^*$ is the $k^{\mathrm{th}}$ sampling instants of the optimal band and $u_i$ is the $i^{\mathrm{th}}$ sampling instant the suboptimal band. Note that since the rewards are bounded in $[0,1]$ and the exploration function is increasing and unbounded, there always exists a time instant when the index of any given band will be the largest (and hence sensed). Consequently both bands will be sensed infinitely many times, i.e., $m^* \to \infty$ and $m \to \infty$.

Take the event

$$\bigg\{\underbrace{\frac{1}{m^*}\sum_{k=1}^{m^*}x^*(u_k^*) - \mu^* + \frac{\epsilon\Delta}{2}}_{A} + \underbrace{\frac{1}{m}\sum_{i=1}^{m}\mu - x(u_i) + \frac{\epsilon\Delta}{2}}_{B} \leq 0,\ \mathrm{i.o.}\bigg\}.$$

Now we can notice that in order for $A + B$ to be negative at least one of $A$ or $B$ have to be negative. Hence, we get that

$$P\{A+B \leq 0, \text{ i.o.}\} \leq P\{A \leq 0, \text{i.o}\} + P\{B \leq 0, \text{i.o}\}$$

$$= P\left\{\frac{1}{m^*}\sum_{k=1}^{m^*}(x^*(u_k^*) - \mu^*) \leq \frac{-\epsilon\Delta}{2}, \text{ i.o.}\right\}$$

$$+ P\left\{\frac{1}{m}\sum_{i=1}^{m}(\mu - x(u_i)) \leq \frac{-\epsilon\Delta}{2}, \text{ i.o.}\right\}$$

$$\leq P\left\{\left|\frac{1}{m^*}\sum_{k=1}^{m^*}(x^*(u_k^*) - \mu^*)\right| \geq \frac{\epsilon\Delta}{2}, \text{ i.o.}\right\}$$

$$+ P\left\{\left|\frac{1}{m}\sum_{i=1}^{m}(\mu - x(u_i))\right| \geq \frac{\epsilon\Delta}{2}, \text{ i.o.}\right\},$$

The last inequality is due to the fact that for any (real) random variable $Y$, $P\{Y < -y\} \leq P\{|Y| > y\}$.

Since $x(n)$ and $x^*(n)$ have finite variances we notice using Lemma 1 that,

$$P\left\{\left|\frac{1}{m^*}\sum_{i=1}^{m^*}(x^*(u_i^*) - \mu^*)\right| \geq \frac{\epsilon\Delta}{2}, \text{ i.o.}\right\} = 0$$

$$P\left\{\left|\frac{1}{m}\sum_{k=1}^{m}(\mu - x(u_k))\right| \geq \frac{\epsilon\Delta}{2}, \text{ i.o.}\right\} = 0.$$

Hence we conclude that

$$P\left\{\bar{x}^*(n) - \bar{x}(n) \leq (1-\epsilon)\Delta, \text{ i.o.}\right\} = 0. \quad (16)$$

In other words, for any $\epsilon > 0$ with probability 1 there exists a time instant $n_0$ when the sample average of the optimal band will be at least $(1-\epsilon)\Delta$ larger than the sample average of the suboptimal band. Consequently, for any two consecutive sensing time instants $z_j$ and $z_{j+1}$ of a suboptimal band, for which $z_{j+1} > z_j > n_0$, the following will hold:

$$z_{j+1} \geq z_j g^{-1}\big((1-\epsilon)\Delta\big). \quad (17)$$

Since $g^{-1}((1-\epsilon)\Delta) > 1$ the difference between the sensing time instants of the suboptimal band will asymptotically grow exponentially. When the time difference between two consecutive sensing time instants increases exponentially the number of sensings must grow logarithmically and hence the expected number of suboptimal sensings is

$$E[M(n)] = O(\ln(n)). \quad (18)$$

## APPENDIX B
## PROOF OF THEOREM 2

In order to prove that the policy in Fig. 4 has logarithmic weak regret with Markovian rewards one needs to show, similarly as in the independent rewards case, that with probability 1 there exists a time instant after which the sample mean of the rewards of the optimal band is always greater than that of the suboptimal band. In the proof for independent rewards in appendix A the past sensing time instants did not play a role in showing the convergence of the reward sample means to the true expected reward from a band. This was because the rewards were assumed to be independent from the past. With Markovian rewards, however, the sensing policy, i.e. how the sensing time instant is selected plays a role whether the reward sample means converge to the true stationary mean or not. In order to simplify the notation we have dropped the indexing of the bands and focus only to one of the bands by showing that its reward sample average converges to its true stationary mean almost surely. Rest of the proof follows essentially the same path as the proof for the i.i.d. case in appendix A.

The exploration bonus of a band that is not sensed grows unboundedly so that the index of that band will also grow unboundedly. As a consequence for any band there always exists a future time instant when its index will be the largest one and when it will be sensed again. Since every time a band is selected for sensing and since it is sensed for at least one full regenerative cycle, the number of regenerative cycles spent on sensing each band approaches infinity as the policy runs infinitely long. This in mind we may use the strong law of large numbers for Markov chains to prove the convergence of the reward sample mean to the true stationary expected reward.

Let $r_0$ denote the reward (data rate) that the CR obtains when it senses and accesses the band in idle state and let $r_1$ denote the reward obtained when the band is sensed and accessed in occupied state. The expected reward from the band is then

$$\mu = r_0 P_0 + r_1 P_1, \quad (19)$$

where $P_0$ is the stationary probability of state 0 and $P_1$ the stationary probability of state 1.

Denote the total number of sensings of the band at time instant $n$ as $M(n)$. Assume that $n$ is the time instant at the end of one of the regenerative cycles. Denote the length of the j$^{\text{th}}$ 0-cycle (regenerative cycle starting and ending in state 0) as $T_j^0$. Notice that $T_j^0$'s are i.i.d. with mean $E[T_j^0] = 1/P_0$ (see e.g. [32] Theorem 1.41), where $P_0$ is the stationary probability of state 0. It can be shown that (see e.g. [37], [31] or [32]) $\overline{T^0}(n) \stackrel{\text{a.e.}}{\to} E[T_j^0]$, where $\overline{T^0}(n)$ is the sample average of the lengths of the 0-cycles observed up to time $n$. This result is a consequence of the independence of the lengths of the regenerative cycles and the strong law of large numbers. Then the following will also hold:

$$\overline{T^0}(n)^{-1} \stackrel{\text{a.e.}}{\to} E[T_j^0]^{-1} = P_0 \quad (20)$$

Similarly for the 1-cycles (regenerative cycles starting and ending in state 1) we have $\overline{T^1}(n) \stackrel{\text{a.e.}}{\to} 1/P_1$ and

$$\overline{T^1}(n)^{-1} \stackrel{\text{a.e.}}{\to} P_1, \quad (21)$$

where $\overline{T^1}(n)$ is the sample average of the lengths of the 1-cycles observed until time $n$.

On the other hand the reward sample average of all the sensings up to time $n$ is $\bar{x}(n) = \frac{S(n)}{M(n)}$, where $S(n)$ is the sum of the rewards collected from the band up to time $n$. It is assumed that the last observed reward whenever the SU decides to hop to another band is not counted in the sample average (although naturally the reward is collected). The sample average can be further expressed as

$$\bar{x}(n) = \frac{r_0(V_0^0(n) + V_0^1(n)) + r_1(V_1^1 + V_1^0)}{M(n)}$$

where $V_u^v(n)$ is the number of visits to state $u$ during the observed $v$-cycles up to time $n$. By denoting the total number of sensings during 0-cycles as $M_0(n)$ and the total number of sensings during 1-cycles as $M_1(n)$ we get $M(n) = M_0(n) + M_1(n)$. Since the channel must either in be in state 1 or state 0 it must hold that $M_1(n) = V_0^1(n) + V_1^1(n)$. Similarly, $M_0(n) = V_0^0(n) + V_1^0(n)$. Using these we may further express the sample average as

$$\bar{x}(n) = \frac{r_0 \left( \frac{V_0^0(n)}{M_0(n)} M_0(n) + \frac{V_0^1(n)}{M_1(n)} M_1(n) \right)}{M_0(n) + M_1(n)}$$
$$+ \frac{r_1 \left( \frac{V_1^1(n)}{M_1(n)} M_1(n) + \frac{V_1^0(n)}{M_0(n)} M_0(n) \right)}{M_0(n) + M_1(n)}.$$

Next we notice that

$$\begin{cases} \frac{V_0^0(n)}{M_0(n)} &= \overline{T_0}(n)^{-1} \\ \frac{V_1^1(n)}{M_1(n)} &= \overline{T_1}(n)^{-1} \\ \frac{V_0^1(n)}{M_1(n)} &= \frac{M_1(m) - V_1^1(n)}{M_1(n)} = 1 - \overline{T_1}(n)^{-1} \\ \frac{V_1^0(n)}{M_0(n)} &= \frac{M_0(m) - V_0^0(n)}{M_0(n)} = 1 - \overline{T_0}(n)^{-1}. \end{cases}$$

Substituting these we get

$$\bar{x}(n) = \frac{r_0 \left( \overline{T_0}(n)^{-1} M_0(n) + \left(1 - \overline{T_1}(n)^{-1}\right) M_1(n) \right)}{M_0(n) + M_1(n)}$$
$$+ \frac{r_1 \left( \overline{T_1}(n)^{-1} M_1(n) + \left(1 - \overline{T_0}(n)^{-1}\right) M_0(n) \right)}{M_0(n) + M_1(n)}.$$

Using (20) and (21) we get that

$$\bar{x}(n) \overset{\text{a.c.}}{\to} = r_0 P_0 + r_1(1 - P_0) = \mu,$$

which is equivalent with

$$P\left\{ |\bar{x}(n) - \mu| > \epsilon, \text{i.o.} \right\} = 0. \tag{22}$$

Now again denote the sample average of the optimal band by $\bar{x}^*(n)$ and a suboptimal band by $\bar{x}(n)$ respectively. Using the same steps as in section A $((15)\text{-}(16))$ with i.i.d. rewards and the result of (22) we get that

$$P\left\{ \bar{x}^*(n) - \bar{x}(n) \leq (1-\epsilon)\Delta, \text{i.o.} \right\} = 0. \tag{23}$$

This means that with probability 1 there exists a finite time instant $n_0$ when the sample average of the optimal band is at least by $(1-\epsilon)\Delta$ larger than that of a suboptimal band and will stay larger from there on. Denoting the time instant of the start of the $j^{\text{th}}$ regenerative cycle at a suboptimal band by $z_j$ for which $z_{j+1} > z_j > n_0$, the following will hold:

$$z_{j+1} \geq z_j g^{-1}\big((1-\epsilon)\Delta\big). \tag{24}$$

Since the Markov chains are assumed to be recurrent, all observed regenerative cycles are of finite length with probability 1. This guarantees that the use of regenerative cycles will not make the policy to get stuck sensing a suboptimal band forever. Since in (24) $g^{-1}((1-\epsilon)\Delta) > 1$ the difference between the start of a new regenerative cycle will asymptotically grow exponentially for any suboptimal band. When the time difference between two consecutive regenerative cycles (that are always finite) increases exponentially the number of sensing must grow logarithmically. Hence the expected number of sensings of any suboptimal band is

$$\mathrm{E}[M(n)] = O(\ln(n)). \tag{25}$$

ACKNOWLEDGMENT

Prof. Santosh Venkatesh, University of Pennsylvania is acknowledged for useful discussions.

The authors wish to thank the anonymous reviewers for their constructive comments that have improved the quality of the paper.

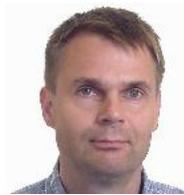

**Visa Koivunen** (IEEE Fellow) received his D.Sc. (EE) degree with honors from the University of Oulu, Dept. of Electrical Engineering. He received the primus doctor (best graduate) award among the doctoral graduates in years 1989-1994. He is a member of Eta Kappa Nu. From 1992 to 1995 he was a visiting researcher at the University of Pennsylvania, Philadelphia, USA. Years 1997 -1999 he was faculty at Tampere UT. Since 1999 he has been a full Professor of Signal Processing at Aalto University (formerly known as Helsinki Univ of Technology) , Finland. He received the Academy professor position (distinguished professor nominated by the Academy of Finland). He is one of the Principal Investigators in SMARAD Center of Excellence in Research nominated by the Academy of Finland. Years 2003-2006 he has been also adjunct full professor at the University of Pennsylvania, Philadelphia, USA. During his sabbatical term year 2007 he was a Visiting Fellow at Princeton University, NJ, USA. He has also been a part-time Visiting Fellow at Nokia Research Center (2006-2012). He spent a sabbatical at Princeton University for the full academic year 2013-2014.

Dr. Koivunen's research interests include statistical, communications, sensor array and multichannel signal processing. He has published about 350 papers in international scientific conferences and journals. He co-authored the papers receiving the best paper award in IEEE PIMRC 2005, EU-SIPCO'2006,EUCAP (European Conference on Antennas and Propagation) 2006 and COCORA 2012. He has been awarded the IEEE Signal Processing Society best paper award for the year 2007 (with J. Eriksson). He served as an associate editor for IEEE Signal Processing Letters, IEEE Transactions on Signal Processing, Signal Processing and Journal of Wireless Communication and Networking. He is co-editor for IEEE JSTSP special issue on Smart Grids. He is a member of editorial board for IEEE Signal Processing Magazine. He has been a member of the IEEE Signal Processing Society technical committees SPCOM-TC and SAMTC. He was the general chair of the IEEE SPAWC conference 2007 conference in Helsinki, Finland June 2007. He is the the Technical Program Chair for the IEEE SPAWC 2015 as well as Array Processing track chair for 2014 Asilomar conference.

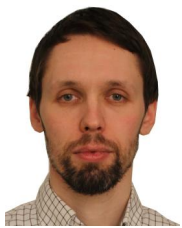

**Jan Oksanen** (S10) received his M.Sc. (with distinction) in 2008 in communications engineering from the Helsinki University of Technology (currently known as Aalto University), Finland. He is currently finalizing his D.Sc. (Tech) at the department of Signal processing and acoustics, Aalto university. Nov. 2011–Nov 2012 he was a visiting student research collaborator at Princeton University, NJ, USA. His research interests include spectrum sensing, cognitive radio and machine learning.